\documentclass[12pt]{article}
\usepackage{amssymb}
\usepackage{amsmath}
\usepackage{stmaryrd}
\catcode`@=11 
\renewcommand{\section}{\@startsection{section}{1}{0pt}{\medskipamount}
{\medskipamount}{\bf}}
\catcode`@=12 
\def\a{\alpha}
\def\b{\beta}

\def\d{\delta}
\def\e{\epsilon}

\def\p{\psi}

\def\m{\mu}
\def\n{\nu}

\def\s{\sigma}

\def\hV{\widehat{V}}
\def\wH{\widetilde{H}}
\def\wD{\widetilde{D}}
\def\wK{\widetilde{K}}
\def\wT{\widetilde{T}}
\def\wJ{\widetilde{J}}
\def\wQ{\widetilde{Q}}
\def\wS{\widetilde{S}}

\newcommand{\cN}{{\cal N}}
\def\be{\begin{equation}}
\def\ee{\end{equation}}
\def\arr{\begin{array}{rll}}
\def\ea{\end{array}}
\def\bea{\begin{eqnarray}}
\def\eea{\end{eqnarray}}
\def\sfrac#1#2{{\textstyle\frac{#1}{#2}}}
\def\sf#1#2{{\textstyle\frac{#1}{#2}}}
\def\ti{{\times}}
\def\ph{\phantom{-}}
\def\ic{{\rm i}}
\def\eu{{\rm e}}

\def\pa{\partial}
\def\>{\rangle}
\def\<{\langle}
\def\+{\dagger}
\def\={\ =\ }
\def\und{\qquad\textrm{and}\qquad}
\begin{document}
\renewcommand{\thefootnote}{\fnsymbol{footnote}}
\begin{titlepage}
\setcounter{page}{0}
\begin{flushright}
ITP--UH--17/07\\
LMP-TPU--9/07 \\
\end{flushright}
\vskip 1cm
\begin{center}
{\LARGE\bf N=4 superconformal Calogero models}
\vskip 2cm
$
\textrm{\Large Anton Galajinsky\ }^{a} ,\quad
\textrm{\Large Olaf Lechtenfeld\ }^{b} ,\quad
\textrm{\Large Kirill Polovnikov\ }^{a}
$
\vskip 0.7cm
${}^{a}$ {\it
Laboratory of Mathematical Physics, Tomsk Polytechnic University, \\
634050 Tomsk, Lenin Ave. 30, Russian Federation} \\
{Emails: galajin, kir @mph.phtd.tpu.edu.ru}
\vskip 0.4cm
${}^{b}$ {\it
Institut f\"ur Theoretische Physik, Leibniz Universit\"at Hannover,\\
Appelstrasse 2, D-30167 Hannover, Germany} \\
{Email: lechtenf@itp.uni-hannover.de}
\vskip 0.2cm
\end{center}
\vskip 2cm
\begin{abstract} \noindent
We continue the research initiated in hep-th/0607215 and apply our
method of conformal automorphisms to generate various $\cN{=}4$
superconformal quantum many-body systems on the real line from a set 
of decoupled particles extended by fermionic degrees of freedom. The
$su(1,1|2)$ invariant models are governed by two scalar potentials 
obeying a system of nonlinear partial differential equations which 
generalizes the Witten-Dijkgraaf-Verlinde-Verlinde equations. 
As an application, the $\cN{=}4$ superconformal extension of the 
three-particle ($A$-type) Calogero model generates a unique
$G_2$-type Hamiltonian featuring three-body interactions. 
We fully analyze the $\cN{=}4$ superconformal three- and four-particle 
models based on the root systems of $A_1\oplus G_2$ and $F_4$, respectively.
Beyond Wyllard's solutions we find a list of new models, whose
translational non-invariance of the center-of-mass motion fails to
decouple and extends even to the relative particle motion.
\end{abstract}
\vspace{2cm}
\end{titlepage}
\renewcommand{\thefootnote}{\arabic{footnote}}
\setcounter{footnote}0

\noindent
{\bf 1. Introduction}\\

\noindent Recently conformally invariant models in one dimension
were investigated extensively \cite{town}--\cite{div}. On the one
hand, the interest derives from the AdS/CFT correspondence.
Although there has been considerable progress in understanding the
AdS/CFT duality \cite{adc/cft}, nontrivial examples of
AdS${}_2$/CFT${}_1$ correspondence are unknown. On the other hand,
the conformal group SO$(2,d{-}1)$ is the isometry group of anti de
Sitter space AdS${}_d$. Since anti de Sitter space describes the
near-horizon geometry of a wide class of extreme black holes (for
a review see e.g. \cite{moh}), it was conjectured \cite{claus,
gibb} that the study of conformally invariant models in $d{=}1$
yields new insight into the quantum mechanics of black holes. This
idea was pushed further in a series of papers
\cite{mich}--\cite{str3}, where some conformal mechanics on
black-hole moduli spaces in $d{=}4$ and $d{=}5$ was constructed
and investigated.

Particularly appealing in this context seems a proposal in
\cite{gibb} that an $\cN{=}4$ superconformal extension of the
Calogero model \cite{calo} might provide a microscopic description
of the extreme Reissner-Nordstr\"om black hole near the horizon.
It should be stressed, however, that the Calogero model, which
describes a pair-wise interaction of $n$ identical particles on
the real line, is not the only multi-particle exactly solvable
conformal mechanics available in $d{=}1$. More complicated systems
describing three-particle and four-particle interactions were
studied in \cite{wolf}--\cite{ruehl}. Since in the context of
\cite{gibb} it is the structure of the conformal algebra which
matters, a priori any multi-particle $\cN{=}4$ superconformal
mechanics seems to be a good starting point. A classification of
(off-shell) $d{=}1$ supermultiplets is interesting in its own
right because of features absent in higher dimensions (see
e.g.~\cite{ils}). In this connection the construction of
multi-particle $\cN{=}4$ superconformal models is relevant for
possible couplings of $d{=}1$, $\cN{=}4$ supermultiplets.

Several attempts have been made to construct an $\cN{=}4$
superconformal extension of the Calogero model
\cite{wyl}--\cite{bgl}. In \cite{wyl} conditions for 
$su(1,1|2)$ invariance were formulated, and some solutions were
presented. In \cite{bgk} the problem was solved for a
complexification of the Calogero model. In \cite{gal,bgl} the
construction of an $\cN{=}4$ superconformal Calogero model was
reduced to solving a system of nonlinear partial differential
equations, which generalizes the
Witten-Dijkgraaf-Verlinde-Verlinde equation known from
two-dimensional topological field theory \cite{w,dvv}. However,
beyond the two-particle case only partial results were obtained.

In the present work we continue the research initiated in
\cite{glp} and apply the method of unitary transformations to
generate various $su(1,1|2)$ invariant quantum many-body systems,
including an $\cN{=}4$ superconformal extension of the Calogero
model. In section~2 we discuss a specific unitary transformation,
which maps a generic conformally invariant model of $n$ identical
particles on the real line to a set of decoupled particles, with
the interaction being pushed into a nonlocal conformal boost
generator. In this description, an $\cN{=}4$ supersymmetric
extension is straightforward to construct as we demonstrate in
section~3. Both the conformal boost generator and its superpartner
are nonlocal in this picture. The inverse transformation then 
provides us with the interacting Hamiltonian. The closure of the 
superconformal algebra poses constraints on the interaction, 
which are detailed and partially solved in section~4. 
Our superconformal models are governed by two scalar potentials 
obeying certain homogeneity conditions and the
Witten-Dijkgraaf-Verlinde-Verlinde-type equations of~\cite{gal,bgl}. 
Explicit three- and four-particle solutions to these ``structure equations'' 
for the two scalar potentials are discussed in section~5 and found to be based 
on certain root systems. Beyond the models found by Wyllard~\cite{wyl},
we present a list of solutions which break translation invariance
not only for the center-of-mass motion but also for the relative motion. In 
section~6 we summarize our results and discuss possible further developments.
\vspace{1cm}

\noindent
{\bf 2. Conformal mechanics in a free nonlocal representation}\\

\noindent
Let us consider a system of $n$ identical particles on the real line with
a Hamiltonian of the generic form
\be\label{h}
H\=\sfrac{1}{2 m} p_i p_i\ +\ V_B (x^1, \dots, x^n)\ ,
\ee
where $m$ stands for the mass of each particle.
Throughout the paper a summation over repeated indices is understood.
Later, the bosonic potential $V_B$ will get supersymmetrically extended to
a potential~$V$ including $V_B$.

For conformally invariant models the Hamiltonian~$H$ is part of
the $so(1,2)$ conformal algebra 
\be\label{al} 
[D,H]\=-\ic\hbar H\ ,\quad 
[H,K]\=2\ic\hbar D\ ,\quad 
[D,K]\=\ic\hbar K\ , 
\ee 
where $D$ and $K$ are the dilatation and conformal boost generators,
respectively. Their realization in term of coordinates and
momenta, subject to 
\be 
[x^i, p_j]\=\ic \hbar {\d_j}^i \ , 
\ee
reads 
\be 
D\=-\sfrac{1}{4} (x^i p_i +p_i x^i) \= D_0 \und 
K\= \sfrac{m}{2} x^i x^i \= K_0 \ , 
\ee 
where the $0$ subscript
indicates the generators in the free model ($V_B{=}0$). The first
relation in (\ref{al}) restricts the potential via \be\label{ucl}
(x^i \partial_i +2)\,V_B \=0\ , \ee meaning that $V_B$ must be
homogeneous of degree $-2$ for the model to be conformally
invariant. In this paper we assume this to be the case. Two simple
solutions to~(\ref{ucl}) are the free model of $n$ non-interacting
particles, 
\be 
V_B\= 0 \qquad\longrightarrow\qquad 
H_0 \= \sfrac{1}{2 m} p_i p_i \ , 
\ee 
and the Calogero model of $n$ particles interacting through 
an inverse-square pair potential,
\be 
V_B \= \sum_{i<j} \sfrac{g^2}{(x^i-x^j)^2} \qquad\longrightarrow\qquad 
H \= H_0 \ +\ V_B\ . 
\ee

As the next step we study the behavior of a generic conformal
multi-particle mechanics under a judiciously chosen
conformal-algebra automorphism. 
Given the particular $so(1,2)$ element 
\be\label{a} 
A \= \a H -2 D + \sfrac1{\a} K
\ee 
for a real parameter~$\a$, let us consider the unitary transformation 
\be\label{bak} 
T \quad\longmapsto\quad T' \= 
\eu^{\frac{\ic}{\hbar} A}\,T\,\eu^{-\frac{\ic}{\hbar} A} 
\ee 
on the $so(1,2)$ generators:\footnote{ 
$A$ was chosen such that the Baker-Haussdorff series 
in (\ref{bak}) terminates at the third step~\cite{glp}.} 
\bea
H\quad\longmapsto\quad
H' &=& \qquad\qquad\qquad \sfrac{1}{\a^2} K \ ,\\[8pt]
D\quad\longmapsto\quad
D' &=& \qquad\quad\ -D+\sfrac{2}{\a} K \ ,\\[10pt]
K\quad\longmapsto\quad
K' &=& \a^2 H-4\a D+4K \ .
\eea

Notice that in the previous consideration it is only the structure of
the conformal algebra that matters. Therefore, an analogous map exists
for the free theory defined by $(H_0,D_0,K_0)$:
\be
T_0 \quad\longmapsto\quad T'_0 \=
\eu^{\frac{\ic}{\hbar} A_0}\,T_0\,\eu^{-\frac{\ic}{\hbar} A_0}
\qquad\textrm{with}\quad
A_0 \= \a H_0 -2 D_0 + \sfrac1{\a} K_0 \ .
\ee
This suggests the idea to combine the $A$-map with inverse $A_0$-map
to link $H$ and $H_0$ in the following scheme:
\be\label{tilded}
\begin{aligned}
(H,D,K) \ \qquad\buildrel{A}\over{\longmapsto}\qquad\quad
(H',D',K') \qquad\=\qquad (H'_0,D'_0,K'_0{+}\a^2V_B) \\[8pt]
(\wH,\wD,\wK) \quad\; :=\,\quad (H_0,D_0,K_0{+}\a^2\hV_B)
\qquad\buildrel{-A_0}\over{\longmapsfrom}\qquad\!(H'_0,D'_0,K'_0{+}\a^2V_B)
\end{aligned}
\ee
with the abbreviation
\be
\hV_B\=\eu^{-\frac{\ic}{\hbar} A_0}\,V_B\,\eu^{\frac{\ic}{\hbar} A_0} 
\qquad\textrm{in}\qquad \wK \= K_0 + \a^2\hV_B\ .
\ee
We remark that the dimensionful parameter~$\a$
simply takes care of the different dimensionalities of the $so(1,2)$
generators and drops out of the final results as was shown in~\cite{glp}.
For the remainder of the paper we set $m=1$.

Thus, with the help of a unitary operation one can
transform a generic multi-particle conformal mechanics (\ref{h}), (\ref{ucl})
into a one describing a system of non-interacting particles.
A peculiar feature of this correspondence is that the generator of special
conformal transformations $\wK$ is {\it nonlocal\/} and
effectively hides the interaction potential. In fact, the interaction has
disappeared in the Hamiltonian~$\wH$ but resurfaced in a nonlocal contribution
to the conformal boost~$\wK$. Hence, the price paid for the simplification of
the dynamics is a nonlocal realization of the full conformal algebra in the
Hilbert space of the quantized conformal mechanics.

As an example, let us consider
the conformal Calogero model describing the inverse-square pair-wise
interaction of $n$ identical particles of unit mass on the real line,
\be\label{calodg}
V_B=\sum_{i<j}^n~ \frac{g^2}{{(x^i{-}x^j)}^2}\ ,
\ee
where $g$ is the coupling constant. 
For this model, a map of $H$ to $H_0$ similar to ours was constructed
in~\cite{pol}. However, the entire $so(1,2)$ algebra was not examined,
and the nonlocal structure present in $\wK$ was not revealed there.
The quantum mechanical scattering analysis of the conformal Calogero
model was accomplished in~\cite{poli}, where it was argued that 
the particles merely exchange their asymptotic momenta without altering 
their values. The asymptotic wave function only picks up an energy-independent
phase factor through the scattering process. Since the conformal Calogero 
particles are indistinguishable, their physics is that of $n$ free bosons. 
Thus, the general consideration presented above is in agreement 
with~\cite{poli,suth}.
\vspace{1cm}

\noindent
{\bf 3. \cN=4 superconformal extension}\\

\noindent
The unitary map of a generic multi-particle conformal mechanics to
a set of decoupled particles considered in the previous section offers 
a novel way to constructing superconformal extensions.
In our setting this amounts to properly adding fermionic degrees of freedom
to a free system and modifying the nonlocal boost generator $\wK$ so as to
close an $\cN$-extended superconformal algebra. Application of the inverse
unitary transformation to the set of free superparticles then produces 
a desired superconformal extension of the original interacting conformal 
mechanics. In this section we discuss the corresponding algebraic framework.

The bosonic sector of the $\cN{=}4$ superconformal algebra $su(1,1|2)$
includes two subalgebras. Along with $so(1,2)$ considered in the previous 
section one also finds the $su(2)$ R-symmetry subalgebra generated by 
$J_a$ with $a=1,2,3$. The fermionic sector is exhausted by the $su(2)$ doublet 
supersymmetry generators $Q_\a$ and ${\bar Q}^\a$ as well as their 
superconformal partners $S_\a$ and ${\bar S}^\a$, with $\a=1,2$,
subject to the hermiticity relations 
\be
{(Q_\a)}^{\dagger}\={\bar Q}^\a \und {(S_\a)}^{\dagger}\={\bar S}^\a\ .
\ee
The bosonic generators are hermitian. The non-vanishing (anti)commutation 
relations in our superconformal algebra read\footnote{
$\s_1$, $\s_2$ and $\s_3$ denote the Pauli matrices.}
\begin{align}\label{algebra}
&
[D,H] \= -\ic \hbar\, H\ , && 
[H,K] \= 2\ic \hbar\, D\ ,
\nonumber\\[4pt]
&
[D,K] \= +\ic \hbar\, K\ , && 
[J_a,J_b] \= \ic \hbar\, \epsilon_{abc} J_c\ ,
\nonumber\\[2pt]
&
\{ Q_\a, \bar Q^\b \} \= 2\hbar\, H {\d_\a}^\b\ , &&
\{ Q_\a, \bar S^\b \} \=
+2\ic\hbar\,{{(\s_a)}_\a}^\b J_a-2\hbar\,D{\d_\a}^\b-\ic\hbar\,C{\d_\a}^\b\ ,
\nonumber\\[2pt]
&
\{ S_\a\,,\, \bar S^\b \} \= 2\hbar\, K {\d_\a}^\b\ , &&
\{ \bar Q^\a, S_\b \} \=
-2\ic\hbar\,{{(\s_a)}_\b}^\a J_a-2\hbar\,D{\d_\b}^\a+\ic\hbar\,C{\d_\b}^\a\ ,
\nonumber
\end{align}
\begin{align}
& 
[D,Q_\a] \= -\sfrac{1}{2} \ic\hbar\, Q_\a\ , && 
[D,S_\a] \= +\sfrac{1}{2} \ic \hbar\, S_\a\ ,
\nonumber\\[4pt]
&
[K,Q_\a] \= +\ic \hbar\, S_\a\ , && 
[H,S_\a] \= -\ic \hbar\, Q_\a\ ,
\nonumber\\[2pt]
&
[J_a,Q_\a] \= -\sfrac{1}{2} \hbar\, {{(\s_a)}_\a}^\b Q_\b\ , && 
[J_a,S_\a] \= -\sfrac{1}{2} \hbar\, {{(\s_a)}_\a}^\b S_\b\ ,
\nonumber\\[4pt]
& 
[D,\bar Q^\a] \= -\sfrac{1}{2} \ic \hbar\, \bar Q^\a\ , && 
[D,\bar S^\a] \= +\sfrac{1}{2} \ic \hbar\, \bar S^\a\ ,
\nonumber\\[4pt]
& 
[K,\bar Q^\a] \= +\ic \hbar\, \bar S^\a\ , && 
[H,\bar S^\a] \= -\ic \hbar\, \bar Q^\a\ ,
\nonumber\\[2pt]
&
[J_a,\bar Q^\a] \= \sfrac{1}{2} \hbar\, \bar Q^\b {{(\s_a)}_\b}^\a\ , && 
[J_a,\bar S^\a] \= \sfrac{1}{2} \hbar\, \bar S^\b {{(\s_a)}_\b}^\a\ .
\end{align}
Here $\e_{123}=1$, and $C$ stands for the central charge.

Following the same strategy as in the previous section, 
we employ the conformal automorphism~(\ref{bak}) and its free inverse as 
indicated in~(\ref{tilded}), with $A$ being of the same form as in~(\ref{a}).
It is very plausible that the new (tilded) generators differ from the free
ones only in the instances of $K$, $S_\a$ and $\bar S^\a$, so we write
(omitting the complex conjugates and suppressing the indices)
\be\label{stilded}
\begin{aligned}
H \qquad&\longmapsto\qquad \wH \= H_0 \ , \\[2pt]
D \qquad&\longmapsto\qquad \wD \= D_0 \ , \\[2pt]
K \qquad&\longmapsto\qquad \wK \= K_0 + \a^2\hV \ , \\[2pt]
Q \qquad&\longmapsto\qquad \wQ \= Q_0 \ , \\[2pt]
S \qquad&\longmapsto\qquad \wS \= 
S_0 - \a\sfrac{\ic}{\hbar}\widehat{[S_0,V]}\ , \\[2pt]
J \qquad&\longmapsto\qquad \wJ \= J_0 \ ,
\end{aligned}
\ee
where the correction to $S_0$ is determined from the form of $\wK$ through
the $[\wK,\wQ]$ commutator in~(\ref{algebra}), and we again use the notation
\be
\widehat{T}\=\eu^{-\frac{\ic}{\hbar} A_0}\,T\,\eu^{\frac{\ic}{\hbar} A_0}\ .
\ee
Note that we have written $V$ instead of $V_B$,
anticipating fermionic and quantum contributions to the Hamiltonian
\be\label{Hcorr}
H \= H_0\ +\ V \qquad\textrm{with}\qquad
V \= V_B + V_F + O(\hbar)\ .
\ee
Given $V$, the $[H,S]$ and $[H,\bar S]$ commutators in~(\ref{algebra})
enforce an interacting part for the supersymmetry generators,
\be\label{Qcorr}
Q_\a \= Q_{0\a} -\sfrac{\ic}{\hbar}\,[S_{0\a},V] \und
\bar Q^\a \= \bar Q^\a_0 - \sfrac{\ic}{\hbar}\,[\bar S^\a_0,V]\ ,
\ee
while all other generators $T$ remain free, i.e.
\be
D\=D_0\ ,\qquad K\=K_0\ ,\qquad S\=S_0 \und J\=J_0\ .
\ee
This is the result of inverting the map (\ref{stilded}) to return from
the tilded generators~$\wT$ to the original ones~$T$.
We shall, however, use the tilded generators~(\ref{stilded}) to find
the form of~$V$.

For a mechanical realization of the $su(1,1|2)$ superalgebra, 
one introduces fermionic degrees of freedom represented by the
operators $\psi^i_\a$ and $\bar\psi^{i\a}$, with $i=1,\dots,n$ and $\a=1,2$, 
which are hermitian conjugates of each other and obey the anti-commutation 
relations\footnote{
Spinor indices are raised and lowered with the invariant
tensor $\e^{\a\b}$ and its inverse $\e_{\a\b}$, where $\e^{12}=1$.
}
\be
\{\p^i_\a, \p^j_\b \}\=0\ , \qquad 
\{ {\bar\p}^{i\a}, {\bar\p}^{j\b} \}\=0\ , \qquad
\{\p^i_\a, {\bar\p}^{j\b} \}\=\hbar\,{\d_\a}^\b \d^{ij}\ .
\ee
In the extended space it is easy to construct the free fermionic generators 
associated with the free Hamiltonian $H_0=\frac{1}{2}p_ip_i$, 
namely (for $m{=}1$)
\be\label{QSfree}
{Q_0}_\a\=p_i \p^i_\a\ , \qquad 
\bar Q_0^\a\=p_i \bar\p^{i\a} \und
{S_0}_\a\=x^i \p^i_\a\ , \qquad 
\bar S_0^\a\=x^i \bar\p^{i\a}\ ,
\ee
as well as $su(2)$ generators
\be\label{Jfree}
{J_0}_a \= \sfrac{1}{2} \bar\p^{i\a} {{(\s_a)}_\a}^\b \p^i_\b\ .
\ee
Notice that these are automatically Weyl-ordered.
The free dilatation and conformal boost operators maintain their bosonic form
\be\label{DKfree}
D_0\= -\sfrac{1}{4}(x^i p_i +p_i x^i) \und K_0\= \sfrac12 x^i x^i \ .
\ee
In contrast to the bosonic case, the free generators~$T_0$ fail to satisfy
the full algebra~(\ref{algebra}). Even for $C{=}0$, the $\{Q,\bar S\}$ and
$\{\bar Q,S\}$ anticommutators require corrections cubic in the fermions,
which we can restrict to $Q$ and~$\bar Q$ as in~(\ref{Qcorr}). Dimensional
analysis reveals that the coefficients of these cubic terms have a dimension
of length${}^{-1}$ and thus cannot be constants. It follows further that $H$ 
contains quadratic and quartic fermionic terms, which are collected
in $V_F$ in~(\ref{Hcorr}). Hence, even for $V_B{=}0$ there does not exist 
a free mechanical representation of the algebra~(\ref{algebra}). 

The generators $\wK$, $\wS$ and $\widetilde{\bar S}$ are nonlocal.
Substituting their form~(\ref{stilded}) into the superconformal algebra 
(\ref{algebra}) one gets a set of restrictions on the form of 
the operator $V$:
\be\label{restr}
\begin{aligned}
&
[K_0,V]\=0\ ,\qquad
[D_0,V]\=-\ic \hbar\,V\ , \qquad  
[J_{0a},V]\=0\ ,
\\[6pt]
&
\{ S_{0\a},[S_{0\b},V]\}\=\hbar^2 \p^i_\a \p^i_\b\ , \qquad 
\{ {\bar S_0}^\a,[{\bar S_0}^\b,V] \}\=\hbar^2 {\bar\p}^{i\a}{\bar\p}^{i\b}\ ,
\\[6pt]
&
\{S_{0\a},[{\bar S_0}^\b,V]\}\=+2\hbar^2 {{(\s_a)}_\a}^\b J_{0a} +
\sfrac{1}{2}\hbar^2(\p^i_\a \bar\p^{i\b}{-}\bar\p^{i\b} \p^i_\a) -
\hbar^2 C {\d_\a}^\b\ ,
\\[6pt]
&
\{{\bar S_0}^\a,[S_{0\b},V]\}\=-2\hbar^2 {{(\s_a)}_\b}^\a J_{0a} -
\sfrac{1}{2}\hbar^2(\p^i_\b \bar\p^{i\a}{-}\bar\p^{i\a} \p^i_\b) + 
\hbar^2 C {\d_\b}^\a\ ,
\\[6pt]
&
\{ [S_{0\a},V],[S_{0\b},V]\}+
\ic\hbar \{Q_{0\a},[S_{0\b},V]\}+
\ic\hbar \{Q_{0\b},[S_{0\a},V]\}\=0\ ,
\\[6pt]
&
\{ [\bar S_0^\a\,,\,V],[\bar S_0^\b\,,\,V]\}+
\ic\hbar \{\bar Q_0^\a\,,\,[\bar S_0^\b\,,\,V]\}+
\ic\hbar \{\bar Q_0^\b\,,\,[\bar S_0^\a\,,\,V]\}\=0\ ,
\\[6pt]
&
\{ [S_{0\a},V],[{\bar S_0}^\b,V]\}+
\ic \hbar \{Q_{0\a},[{\bar S_0}^\b,V]\}+
\ic \hbar \{{\bar Q_0}^\b,[S_{0\a},V]\}+
2\hbar^3 V {\d_\a}^\b\=0\ ,
\\[6pt]
&
\ic \hbar [Q_{0\a},V]-[H_0+V,[S_{0\a},V]]\=0\ , \qquad
\ic \hbar [{\bar Q_0}^\a,V]-[H_0+V,[{\bar S_0}^\a,V]]\=0\ .
\end{aligned}
\ee
Notice that the vanishing (anti)commutators discarded in (\ref{algebra})
should be taken into account as they also give constraints on $V$. 
For obtaining (\ref{restr}) the following identities are helpful:
\be\label{ident}
\begin{aligned}
&
\widehat{Q}_{0\a}\=2\,Q_{0\a}+\sfrac{1}{\a}S_{0\a}\ , \qquad\qquad &&
\widehat{\bar Q}{}_0^\a\=2\,\bar Q_0^\a+\sfrac{1}{\a} \bar S_0^\a\ ,
\\[2pt]
&
\widehat{S}_{0\a}\=-\a\,Q_{0\a}\ ,&&
\widehat{\bar S}{}_0^\a\=-\a\,\bar Q_0^\a\ ,
\\[2pt]
&
S_{0\a}\=\a\,\widehat{Q}_{0\a}+2\,\widehat{S}_{0\a}\ ,&&
\bar S_0^\a\=\a\,\widehat{\bar Q}{}_0^\a+2\,\widehat{\bar S}{}_0^\a\ .
\end{aligned}
\ee
\vspace{1cm}

\noindent
{\bf 4. The structure equations}\\

\noindent
Let us discuss the structure of solutions to the constraints (\ref{restr}).
The first line in (\ref{restr}) implies that the potential 
$V=V_B+V_F+O(\hbar)$ transforms as a scalar under SU(2) and is a 
degree $-2$ homogeneous function of the $x^i$.
It is straightforward to check that an ansatz for $V_F$ quadratic in 
$\p^i$ and $\bar\p^i$ fails to solve (\ref{restr}). 
This is in contrast with $\cN{=}2$ superconformal extensions~\cite{glp,fm}.
Thus it seems natural to try a general ansatz quartic in the fermionic 
coordinates,\footnote{
The classical consideration in \cite{bgl} implies that (\ref{ans}) is indeed
the most general quartic ansatz compatible with the $\cN{=}4$ superconformal 
algebra.}
\be\label{ans}
V\=V_B(x)\ +\ \hbar O_1(x)\ +\ \hbar^2 O_2(x)\ +\
M_{ij}(x) \langle \p^i_\a {\bar\p}^{j\a} \rangle\ +\ 
\sfrac14 L_{ijkl}(x) \langle\p^i_\a\p^{j\a}\bar\p^{k\b}\bar\p^l_\b\rangle\ ,
\ee
with completely symmetric unknown functions $M_{ij}$ and $L_{ijkl}$.
Here, the symbol $\langle\dots\rangle$ stands for symmetric (or Weyl) ordering
(for our conventions see appendix A),
and the contributions $\hbar O_1(x)$ and $\hbar^2 O_2(x)$ were included to 
account for the ordering ambiguity present in the fermionic sector.
The argument $x$ indicates dependence on $\{x^1,\ldots,x^n\}$.

Introducing the notations
\be\label{wy}
L_{ijkl}\,x^l \ =:\ -W_{ijk} \und M_{ij}\,x^j \ =:\ Y_i
\ee
and substituting the ansatz (\ref{ans}) into the constraints (\ref{restr}),
one obtains the following system of partial differential and algebraic
``structure equations'',
\bea
&&
L_{ijkl}\=\pa_i W_{jkl}\=\pa_j W_{ikl}\ , \qquad\qquad 
M_{ij} \=-\pa_i Y_j   \=-\pa_j Y_i\ ,
\label{lw1} \\[6pt]
&&
x^i W_{ijk}\=-\d_{jk}\ , \qquad\qquad\qquad\qquad\ \
x^i Y_i    \=-C\ ,
\label{lw2} \\[6pt]
&&
M_{ij}+W_{ijk}Y_k\=0\ , \qquad\qquad\qquad\quad\,
W_{ikp} W_{jlp} \= W_{jkp} W_{ilp}\ ,
\label{lw3}
\eea
as well as a boundary condition on $Y_i$,
\be\label{bc}
\sfrac12\,Y_i Y_i \= V_B\ .
\ee
Besides, one determines the quantum corrections as
\be\label{qcor}
O_1\=0 \und O_2\=\sfrac18\,W_{ijk} W_{ijk}\ .
\ee
In contrast to $\cN{=}2$ superconformal models, here the algebra requires
a nontrivial quantum correction. The explicit derivation of
(\ref{lw1})--(\ref{qcor}) is tedious and most efficiently achieved using
reordering relations given in appendix A.

Taking into account that $W_{ijk}$ is a completely symmetric function, 
from~(\ref{lw1}) one finds
\be\label{pot}
\begin{aligned}
W_{ijk} \= \pa_i\pa_j\pa_k F \,\qquad&\Leftrightarrow\qquad &&
L_{ijkl}\= \pa_i\pa_j\pa_k\pa_l F \ ,\\[6pt]
Y_i \= \pa_i U \qquad\qquad&\Leftrightarrow &&\ 
M_{ij} \= -\pa_i\pa_j U\ ,
\end{aligned}
\ee
with two scalar potentials $F(x)$ and $U(x)$ to be determined.
Thus, these scalars govern the $\cN{=}4$ superconformal extension 
and obey the following system of nonlinear partial differential equations,
\bea
&&
(\pa_i\pa_k\pa_p F)(\pa_j\pa_l\pa_p F)\=
(\pa_j\pa_k\pa_p F)(\pa_i\pa_l\pa_p F)\quad,\qquad\qquad
x^i \partial_i \partial_j \partial_k F\=-\d_{jk}\ ,
\label{w1}
\\[6pt]
&&
\pa_i\pa_j U -(\pa_i\pa_j\pa_k F)\,\pa_k U\=0\ ,\qquad
\sfrac{1}{2} (\pa_i U)(\pa_i U)\=V_B\ , \qquad
x^i \pa_i U\=-C\ .
\label{w2}
\eea
Notice that $F$ is defined modulo a quadratic polynomial while $U$ is defined
up to a constant. 

Wyllard~\cite{wyl} obtained equivalent equations, but employed a different 
fermionic ordering. In contrast to his equations, $\hbar$ does not appear
in (\ref{w1}) or~(\ref{w2}), since our Weyl-ordering prescription matches
smoothly to the classical limit.
For the classical Calogero model similar equations were 
discussed in~\cite{bgl}.
 
The right-most equations in (\ref{w1}) and~(\ref{w2}) are inhomogeneous 
with constants $\d_{jk}$ and $C$ (the central charge) on the right-hand side 
and display an explicit coordinate dependence. 
Furthermore, the second equation in (\ref{w1}) can be integrated twice
to obtain
\be\label{w3}
x^i \pa_i F -2F +\sfrac{1}{2} x^i x^i \=0\ ,
\ee
where we used the freedom in the definition of $F$ to put the integration 
constants -- a linear function on the right-hand side -- to zero.
It is important to realize that the inhomogeneous term in this integrated
equation does break translation invariance and excludes the trivial solution
$F=0$ equivalent to a homogeneous quadratic polynomial. This effect 
is absent in $\cN{=}2$ superconformal models, where the four-fermion
potential term is not needed and, hence, $F$ does not appear~\cite{glp}.
This issue is also discussed in~\cite{wyl}.

To be more explicit, we extract the center-of-mass dynamics by splitting
\be\label{comsplit}
F \= F_{\rm com}(X) + F_{\rm rel}(x) \und
U \= U_{\rm com}(X) + U_{\rm rel}(x)
\ee
with the center-of-mass coordinate $X:=\frac1n\sum_{i=1}^n x^i$.
If the {\it relative\/} particle motion is translation invariant
(which need not be the case), then
\be
\sum_{i=1}^n \pa_i F_{\rm rel} \= 0 \= \sum_{i=1}^n \pa_i U_{\rm rel}
\ee
and, applying $\sum_i\pa_i$ to (\ref{w3}) and the last equation in~(\ref{w2}),
we readily find
\be
X\,F''_{\rm com} - F'_{\rm com} \= -n\,X \qquad\textrm{but}\qquad
X\,U''_{\rm com} + U'_{\rm com} \= 0 \ ,
\ee
which are solved by
\be\label{comsol}
F_{\rm com} \= -\sfrac{n}{2} X^2\,\ln|nX| + \lambda X^2 + \mu \und
U_{\rm com} \= -g_0\,\ln|nX| + \nu
\ee
with free constants $\lambda$, $\mu$, $\nu$ and $g_0$.
Clearly, in this case we may put to zero $U_{\rm com}$ but not $F_{\rm com}$,
so that for $g_0{=}0$ we end up with a center-of-mass contribution
\be
V_{\rm com} \= \sfrac{\hbar^2}{8n}\,X^{-2} \ +\ 
\sfrac{n}{4}\,X^{-2}\,\< \Psi_\a\Psi^\a\bar\Psi^\b\bar\Psi_\b\>
\qquad\textrm{with}\qquad \Psi_\a\ :=\ \frac1n\sum_{i=1}^n\psi_\a^i\ .
\ee
Hence, one can separate a translation-invariant relative motion from 
the center-of-mass motion, but the latter is non-linear due to an 
$X^{-2}$ potential as enforced by the superconformal algebra~(\ref{algebra}).

Our attack on (\ref{w1}) and~(\ref{w2}) begins with the homogeneity conditions
\be\label{w4}
(x^i\pa_i - 2) F \= -\sfrac12\,x^ix^i \und x^i\pa_i U \= -C\ .
\ee
The most general solution is the sum of a particular solution and the
general solution to the homogeneous equations,
\be
(x^i\pa_i - 2) F_{\rm hom} \= 0 \und x^i\pa_i U_{\rm hom} \= 0\ ,
\ee
which is spanned by the homogeneous functions of degree two and zero,
respectively. For a particular solution to~(\ref{w4}), we make the ansatz
\be\label{Fansatz}
F \= -\sum_{\m=0}^d h_\m\,\sfrac12(z^\m)^2\,\ln|z^\m| \und
U \= -\sum_{\m=0}^d g_\m\,\ln|z^\m|
\ee
with a certain number ($d{+}1$) of linear coordinate combinations
\be
z^\m \= n^\m_i\,x^i \qquad\textrm{beginning with}\qquad 
z^0  \= n\,X \= {\textstyle\sum_i} x^i\ .
\ee
The relative motion is translation invariant if $\sum_i n_i^\m=0$ for $\m{>}0$.
Compatibility with the conditions~(\ref{w4}) directly yields
\be\label{hcond}
\sum_{\m=0}^d h_\m\,n_i^\m n_j^\m \= \d_{ij} \und
\sum_{\m=0}^d g_\m \= C\ .
\ee
The second relation fixes the central charge, and
the first relation amounts to a decomposition of the identity $(\d_{ij})$
into rank-one projectors. It turns out that the $g_\m$ are independent 
free couplings (if not forced to zero) while the $h_\m$ are not.

A {\it minimal\/} solution involves 
$d{+}1=n$ mutually orthogonal vectors~$n^\m$
beginning with $\vec n^0=(1,1,\ldots,1)$ and normalized as
\be\label{minimal}
\vec n^\m \cdot \vec n^\n \ \equiv\ {\textstyle\sum_i} n_i^\m n_i^\n \=
h_\m^{-1} \d^{\m\n}\ .
\ee
{}From (\ref{Fansatz}) we derive
\be\label{YWform}
W_{ijk} \= -\sum_{\m=0}^{n-1} h_\m\,\frac{n_i^\m n_j^\m n_k^\m}{z^\m} \und
Y_i \= -\sum_{\m=0}^{n-1} g_\m\,\frac{n_i^\m}{z^\m}\ ,
\ee
and for the minimal choice~(\ref{minimal}) the bosonic potential becomes
\be\label{minpot}
V_B \= \frac12\sum_{\m=0}^{n-1} \frac{g_\m^2 h_\m^{-1}}{(z^\m)^2} \und
O_2 \= \frac18\sum_{\m=0}^{n-1} \frac{h_\m^{-1}}{(z^\m)^2}\ ,
\ee
which demonstrates that the quantum corrections only renormalize the 
coupling constants,
\be
g_\m^2 \qquad\longmapsto\qquad 
\widetilde{g}_\m^2 \ :=\ g_\m^2\ +\ \sfrac14 \hbar^2 \qquad\forall~\m\ .
\ee

It is instructive to first investigate small values of~$n$.
At $n{=}2$, relative translation invariance demands
$\vec n^0{=}(1,1)$ and $\vec n^1{=}(1,-1)$ 
with $h_0{=}h_1{=}\frac12$, whence
\be
\begin{aligned} &
F_{\rm rel}\=-\sfrac14\,(x^1{-}x^2)^2\,\ln|x^1{-}x^2| \und
U_{\rm rel}\=-g_1\ln|x^1{-}x^2|\ , \\[6pt] &
W_{\cdot\cdot\cdot} \= 
-\frac12\Bigl( \frac{1}{x^1{+}x^2} \pm \frac{1}{x^1{-}x^2} \Bigr) \und
V_B+\hbar^2O_2 \= \frac{\widetilde{g}_0^2}{(x^1{+}x^2)^2}
+ \frac{\widetilde{g}_1^2}{(x^1{-}x^2)^2}\ .
\end{aligned}
\ee
Beyond $n{=}2$, minimal choices are no longer invariant 
modulo sign under all permutations of the positions~$x^i$, but, 
due to the linearity of~(\ref{w4}), 
this can be remedied by finally summing over all permutations. The 
result is, in general, an overcomplete set of $d{+}1>n$ non-orthogonal vectors.
In section~5 we shall find a non-minimal one-parameter set (in~$F$) of
$n{=}3$ solutions to all structure equations for the choice 
\be\label{3vec}
\vec n^0=(1,1,1)\ ,\quad 
\vec n^1=(1,-1,0)\ ,\quad 
\vec n^2=(1,1,-2) \qquad
\textrm{plus three permutations}\ .
\ee
However, a nontrivial $U_{\rm rel}$
based either on $\vec n^1$ or on $\vec n^2$ appears only for two specific 
parameter values. One may recognize here the root system of $A_1\oplus G_2$,
which is the even part of the root system of the Lie superalgebra~$G_3$.
In the same section, we will describe five one-parameter families of
$n{=}4$ solutions based on (parts of) the $F_4$ root system. Here, only
three discrete models have $U_{\rm rel}$ non-vanishing, but for two of these
the relative particle motion is not translation invariant.

In order to discover these and other solutions to the structure equations,
within our ansatz~(\ref{Fansatz})--(\ref{hcond}) it remains to solve 
the two left-most equations in (\ref{w1}) and (\ref{w2}),
\be\label{YW}
\pa_i Y_j - W_{ijk}\,Y_k \= 0 \und W_{ikp}W_{jlp}\=W_{jkp}W_{ilp}
\ee
for $Y_i=\pa_iU$ and $W_{ijk}=\pa_i\pa_j\pa_kF$. This is quite tough because
of their nonlinearity, and we address them in the following section.
Already we notice, however, that the full system of structure equations 
(\ref{w1}) and~(\ref{w2}) can be attacked in two different ways. 
One possibility, pursued in subsection~5.1, is to start with a given 
conformal potential~$V_B$, e.g.~of Calogero form, find a corresponding~$U$, 
hence~$Y$, and then search for a solution~$W$ to (\ref{YW}) before integrating
it to~$F$. Alternatively, as in subsection~5.2, one can take a particular 
solution $F$ of the quadratic relations in~(\ref{YW}), then find 
a solution~$Y$ to the first equation in~(\ref{YW}) and integrate it to~$U$, 
thereby determining~$V_B$ afterwards. The second strategy will yield 
$\cN{=}4$  superconformal models generalizing the Calogero one. Finally, 
any full solution $(Y,W)$ also determines the $su(1,1|2)$ generators as
\be
\begin{aligned}
& Q_\a \= (p_k+\ic Y_k)\,\p_\a^k \ \,+ 
\sfrac{\ic}{2} W_{ijk}\,\<\p^i_\b\,\p^{j\b}\bar\p^k_\a\> \ ,\\[6pt]
& \bar Q^\a \= (p_k-\ic Y_k)\,\bar\p^{k\a} +
\sfrac{\ic}{2} W_{ijk}\,\<\p^{i\a}\bar\p^{j\b}\bar\p^k_\b\>\ ,\\[6pt]
& H\,\;\=\ \sfrac12 p_ip_i\ +\ \sfrac12 Y_iY_i\ +\ 
\sfrac{\hbar^2}{8} W_{ijk}W_{ijk}\ -\ 
\pa_iY_j\,\< \p^i_\a {\bar\p}^{j\a} \>\ +\
\sfrac14 \pa_iW_{jkl}\,\< \p^i_\a\p^{j\a}\bar\p^{k\b}\bar\p^l_\b \>\ ,
\end{aligned}
\ee
while the other generators are of bilinear form given in 
(\ref{QSfree}), (\ref{Jfree}) and~(\ref{DKfree}). 

We conclude the section by observing a resemblance of the quadratic relations
in (\ref{YW}) or (\ref{w1}) to an $n$-parametric potential deformation of an 
$n$-dimensional Fr\"obenius algebra~\cite{dub}, which plays an important role 
in two-dimensional topological field theory~\cite{w,dvv}.
Let us recall that an $n$-dimensional commutative associative algebra~$A$
with unit element~$e$ is called a Fr\"obenius algebra if it is supplied with 
a non-degenerate symmetric bilinear form obeying 
(for a review see e.g.~\cite{dub})
\be\label{frob}
\langle a \cdot b\,,\,c\rangle \= \langle a\,,\,b \cdot c \rangle 
\qquad \forall~ a,b,c \in A\ .
\ee
Choosing a basis $\{e_i\,|\,i=1,\ldots,n\}$ with $e_1=e$, one has
\be
\langle e_i,e_j\rangle\=\eta_{ij} \und
e_i \cdot e_j \={f_{ij}}^k \, e_k\ ,
\ee
where $\eta_{ij}$ is the metric with inverse $\eta^{ij}$ and ${f_{ij}}^k$ 
are the structure constants. The commutativity and associativity of the 
algebra along with~(\ref{frob}) produce the constraints
\be\label{f1}
{f_{ij}}^k\={f_{ji}}^k\ ,\qquad {f_{1i}}^j\={\d_i}^j\ ,\qquad
{f_{ij}}^k \eta_{kl}\={f_{lj}}^k \eta_{ki}\ ,\qquad
{f_{ij}}^k {f_{kl}}^m \={f_{lj}}^k {f_{ki}}^m\ .
\ee
Thus, ${f_{ij}}^k \eta_{kl} = f_{ijl}$ is totally symmetric and subject to 
the quadratic relations above.

An $n$-parametric potential deformation of such a Fr\"obenius algebra is
defined by a set of functions
\be
f_{ijk}(x)\=\partial_i \partial_j \partial_k F(x)
\ee
descending from some scalar potential $F(x)$ with $x=\{x^1,\ldots,x^n\}$.
To qualify as a deformation, these functions must satisfy the relations
\be\label{wit}
f_{1ij}(x)\=\eta_{ij}\ , \qquad \pa_i\eta_{jk}\=0\ , \qquad
\eta^{kn} f_{ijk}(x) f_{lmn}(x)\=\eta^{kn} f_{ljk}(x) f_{imn}(x)\ ,
\ee
which represent nonlinear partial differential equations for~$F(x)$.
In the context of two-dimensional topological field theory,
$F$ is known as the free energy, and (\ref{wit}) is called the 
Witten-Dijkgraaf-Verlinde-Verlinde (WDVV) equation~\cite{w,dvv}.
An interesting link between the WDVV equation and
differential geometry was established in \cite{dub}.
Comparing (\ref{w1}) with (\ref{wit}), we see that our algebra does not 
have a distinguished element serving as a unit element. Instead, the
metric arises from the second equation in~(\ref{w1}) by a contraction
of $f_{ijk}$ with the coordinates~$x^i$.
\vspace{1cm}

\noindent
{\bf 5. Solutions to the structure equations }\\

\noindent
Proving the integrability of the structure equations (\ref{w1}) and (\ref{w2})
is a difficult task. For the WDVV equations this was done rigorously
only for the simpler case of a decomposable Fr\"obenius algebra~\cite{dub}. 
So, instead of trying to find a formal proof, we shall consider a few
explicit examples and outline a simple constructive procedure how 
to integrate the structure equations. 
Finally, we give all solutions of the three- and four-particle cases
which fit in our ansatz~(\ref{Fansatz}) with $A_1\oplus G_2$ and $F_4$ 
positive root vectors, respectively.
\vspace{0.5cm}

\noindent
{\it 5.1. Three-body $\cN{=}4$ superconformal Calogero model}\\
\noindent
In this subsection we construct a particular solution to (\ref{w1}) and 
(\ref{w2}) or, equivalently, (\ref{lw1})--(\ref{bc}),
for the case of three-body Calogero model governed by the potential
\be\label{bounda}
V_B\=\sum_{i<j}^3 \frac{g^2}{{(x^i{-}x^j)}^2}\ ,
\ee
leading to $C{=}3g$ because we sum over three permutations.
It is easy to construct a corresponding $U$ satisfying the second
and third equation in~(\ref{w2}). The general solution reads
\be
U\=-g \sum_{i<j}^3 \ln|x^i{-}x^j| \ +\ L(y,z)\ ,
\ee
where $L$ is an arbitrary function of the ratios
\be
y=\sfrac{x^1}{x^2} \und  z=\sfrac{x^1}{x^3}
\ee
subject to
\be
(\pa_i L)(\pa_i L) \= 
g \sum_{i\neq j} \sfrac{\pa_i L-\pa_j L}{x^i-x^j}\ ,
\ee
and so we may put $L\equiv0$, which we do for simplicity.
Models based on the potential $\ U=-g\sum_{i<j}\ln|x^i{-}x^j|\ $ 
we term `Calogero'.

Next we turn to the WDVV coefficients $W_{ijk}$, 
of which there are ten for $n{=}3$.
The six linear relations in the first equation of~(\ref{lw2}) allow us 
to express the WDVV coefficients in terms of four objects. 
In order to find their explicit form, we integrate~(\ref{w3}) to
\be
F\=-\frac{1}{2} \left( x^i x^i \ln |x^1| - {(x^1)}^2 \Delta(y,z)\right)\ ,
\ee
where $\Delta(y,z)$ is an unknown function to be determined below, 
and we have distinguished the $x^1$ coordinate.
Triple differentiation of $F$ yields
\be\label{wdvv-c}
\begin{aligned}
&
x^1 W_{111}=-1 -\sfrac{1}{y^2} -\sfrac{1}{z^2} +
\Sigma_1 +\Sigma_2 +3\Sigma_3 +3\Sigma_4 \ ,\quad
x^1 W_{123}=y z( \Sigma_3 + \Sigma_4)\ ,\\[2pt]
&
x^1 W_{112}=\sfrac{1}{y} -y \Sigma_1 -y \Sigma_3 -2y \Sigma_4 \ , \quad
x^1 W_{113}=\sfrac{1}{z} -z \Sigma_2 -2 z \Sigma_3 - z \Sigma_4 \ ,\\[2pt]
&
x^1 W_{122}=-1+ y^2 \Sigma_1 + y^2 \Sigma_4 \ , \quad
x^1 W_{133}=-1+ z^2 \Sigma_2 + z^2 \Sigma_3\ ,\\[4pt]
&
x^1 W_{222}=-y^3 \Sigma_1\ , \quad 
x^1 W_{223}=-z y^2 \Sigma_4\ ,\quad
x^1 W_{233}=-y z^2 \Sigma_3\ , \quad 
x^1 W_{333}=-z^3 \Sigma_2\ ,
\end{aligned}
\ee
with four subsidiary functions
\be\label{sigmas}
\begin{aligned}
&
\Sigma_1 \=\frac12 y^3 \frac{{\pa}^3 \Delta}{\pa y^3} +
3 y^2 \frac{{\pa}^2 \Delta}{\pa y^2} +3y \frac{\pa \Delta}{\pa y}\ ,\qquad &&
\Sigma_2 \=\frac12 z^3 \frac{{\pa}^3 \Delta}{\pa z^3} +
3 z^2 \frac{{\pa}^2 \Delta}{\pa z^2} +3z \frac{\pa \Delta}{\pa z}\ ,\\[2pt]
&
\Sigma_3 \=\frac12 y z^2 \frac{{\pa}^3 \Delta}{\pa y \pa z^2} +
y z \frac{{\pa}^2 \Delta}{\pa y \pa z}\ , &&
\Sigma_4 \=\frac12 z y^2 \frac{{\pa}^3 \Delta}{\pa z \pa y^2} +
y z\frac{{\pa}^2 \Delta}{\pa y \pa z}\ .
\end{aligned}
\ee
In order to complete the analysis, we examine the first equation
of~(\ref{lw3}), which couples the two scalar potentials.
It yields six linear algebraic equations for the WDVV coefficients,
but only three are independent. Abbreviating
\be\label{abc}
\begin{aligned}
&
a\=\left(y \partial_2 -\partial_1\right) U\ , &&
b\=\left(z \partial_3 -\partial_1\right) U\ , \quad &&
m\=\left( x^1 \partial_2 \partial_2 +\partial_1 \right) U\ ,\\[2pt]
&
p\=\left( x^1 \partial_3 \partial_3 +\partial_1 \right) U\ , \quad &&
n\=x^1 \partial_2 \partial_3 U\ ,
\end{aligned}
\ee
one finds
\be\label{sigma123}
\Sigma_1\=-\frac{m}{ay^2}-\frac{b}{a}\Sigma_4\ , \qquad
\Sigma_2\=-\frac{p}{bz^2}+\frac{an}{b^2yz}+\frac{a^2}{b^2}\Sigma_4\ , \qquad
\Sigma_3\=-\frac{n}{byz}-\frac{a}{b}\Sigma_4\ .
\ee

In order to fix the last missing coefficient $\Sigma_4$, one is to analyze the
WDVV equations, i.e.~the second relation in~(\ref{lw3}). 
Using the explicit representation (\ref{wdvv-c}) it is
straightforward to verify that among the six nontrivial WDVV equations
at $n{=}3$ only one is independent, namely
\be\label{w-f}
W^{22p}\,W^{33p}\=W^{23p}\,W^{23p}\ .
\ee
With the help of (\ref{sigma123}) this reduces to a linear equation, 
which determines $\Sigma_4$ as
\begin{eqnarray}\label{sigma4}
\Sigma_4\=\frac{1}{18y}\Bigl(\frac{9}{y-z}+\frac{6}{y+z+yz}-
\frac{2}{2y-z-yz}+\frac{4}{2z-y-yz}+\frac{1}{2yz-y-z}\Bigr)\ ,
\end{eqnarray}
and therewith $\Sigma_1$, $\Sigma_2$ and $\Sigma_3$.

The fact that for the three-body problem the WDVV equation (\ref{w-f}) turns 
out to be linear can be understood in a different way. One can extract from 
the WDVV equation linear consequences which, along with other equations in 
(\ref{lw1})--(\ref{bc}), already contain all the information in~(\ref{w-f}).
Indeed, let us differentiate the middle equation in (\ref{w2}),
\be
(\pa_j \pa_i U) (\pa_i U) \=\pa_j V_B\ ,
\ee
and contract the first equation in (\ref{w2}) with $\pa_i U$,
\be
\pa_j V_B\=W_{ijk}(\pa_i U) (\pa_k U)\ .
\ee
Now contracting the WDVV equation with $(\pa_i U)(\pa_j U)$
and taking into account the first equation in (\ref{w2}) one gets the linear
equations
\be\label{suplin}
(\pa_i \pa_k U)( \pa_j \pa_k U)-W_{ijk}\,\pa_k V_B\=0\ .
\ee
It is straightforward to verify that only one component in (\ref{suplin})
is independent and contains just the same information as (\ref{w-f}).

Having fixed the WDVV coefficients algebraically, we are now in a position 
to find the potential $F$.
Substituting (\ref{sigma123}) and (\ref{sigma4}) into (\ref{sigmas}), 
one obtains for the single function $\Delta$ a system of partial 
differential  equations of the Euler type. The standard change of variables
\be
y\=\eu^t \und z\=\eu^s
\ee
turns it into a system of partial differential equations with constant 
coefficients. The latter is readily integrated by conventional means
(see e.g.~\cite{smirnov}) and yields the following free energy,
\be\label{fe}
\begin{aligned}
F(x^1,x^2,x^3) \= 
& -\sfrac16 (x^1{+}x^2{+}x^3)^2 \ln|x^1{+}x^2{+}x^3| \ + \\[6pt]
& -\sfrac14 \sum_{i<j} (x^i{-}x^j)^2 \ln|x^i{-}x^j| \ +\ \sfrac1{36}\!\!
\sum_{i<j\atop i\ne k\ne j} (x^i{+}x^j{-}2x^k)^2 \ln|x^i{+}x^j{-}2x^k|\ ,
\end{aligned}
\ee
revealing the values 
\be
h_0 = \sfrac13\ ,\quad h_1 = \sfrac12\ ,\quad h_2 = -\sfrac{1}{18}
\ee
in the ansatz~(\ref{Fansatz}) for the three types of roots in~(\ref{3vec}).
The relative particle motion is translation invariant.
Note that each sum contains three terms, 
so that the result is totally symmetric in $\{x^1,x^2,x^3\}$. 
Six constants of integration enter a polynomial quadratic in $x$,
which can be discarded since $F$ is defined up to such a polynomial. 
The quantum correction to the Calogero potential finally reads
\be\label{n3qu}
O_2 \= 
\sfrac38\,(x^1{+}x^1{+}x^3)^{-2}\ +\ 
\sfrac14\sum_{i<j} (x^i{-}x^j)^{-2}\ +\
\sfrac1{12} \sum_{i<j\atop i\ne k\ne j} (x^i{+}x^j{-}2x^k)^{-2} \ .
\ee
For the reader's convenience we display the corresponding
WDVV coefficients in appendix~B.

The $\cN{=}4$ superconformal extension of the three-particle Calogero system
produced a unique $G_2$-type integrable model with one free coupling and
particular three-body interactions~\cite{wolf}. Despite the latter, we call 
this a Calogero model because its bosonic classical potential~$V_B$ is just 
the ($A$-type) Calogero one. This terminology differs from the one of
Wyllard~\cite{wyl}, who allowed for three-body interactions in $U$ and $V_B$
from the outset. Our model agrees with his second one-parameter solution.
\vspace{0.5cm}

\noindent
{\it 5.2. A four-body $\cN{=}4$ superconformal model}\\
In this section we consider the second strategy outlined after~(\ref{YW})
and construct a four-body $\cN{=}4$ superconformal model
starting from a solution $F$ to the WDVV equations.
For $n{=}4$ we make the following ansatz for the potential~$F$,
\be\label{n4ansatz}
\begin{aligned}
F(x^1,x^2,x^3,x^4) \= 
&-\sfrac12 h_0\,(x^1{+}x^2{+}x^3{+}x^4)^2 \ln|x^1{+}x^2{+}x^3{+}x^4|\ +\\[6pt]
&-\sfrac12 h_1\!\!\!\sum_{j>i<k<l\atop k\ne j\ne l}
(x^i{+}x^j{-}x^k{-}x^l)^2 \ln|x^i{+}x^j{-}x^k{-}x^l|
\end{aligned}
\ee
where the permutation sum has three terms.
Note that the chosen positive root vectors
\be
\vec n^0=(1,1,1,1)\ ,\quad
\vec n^1=(1,1,-1,-1)\ ,\quad (1,-1,1,-1)\ ,\quad(1,-1,-1,1)
\ee
give translation-invariant relative motion and
form an orthogonal set, i.e.~we look at a minimal model 
with a $A_1{\oplus}A_1{\oplus}A_1{\oplus}A_1$ root system.
Substituting the ansatz into~(\ref{w3}), one learns that
\be\label{ab}
h_0\=h_1\=\sfrac14\ ,
\ee
in agreement with the minimal property $\ h_\m^{-1}=\vec n^\m{\cdot}\vec n^\m$
from~(\ref{minimal}).
For the case at hand one finds twenty WDVV equations, which happen 
to be satisfied identically for the above value of $h_0$ and~$h_1$.

Let us take the corresponding ansatz for~$U$,
\be
U \= 
-g_0\,\ln|x^1{+}x^2{+}x^3{+}x^4|\ -\ g_1 
\!\!\!\sum_{j>i<k<l\atop k\ne j\ne l} \ln|x^i{+}x^j{-}x^k{-}x^l| \ ,
\ee
where $g_0$ and $g_1$ play the role of two independent coupling constants. 
It is straightforward to verify that the first equation in (\ref{w2}) holds 
without imposing any restrictions on the form of the coupling constants.
The last equation in~(\ref{w2}) determines the value of the central charge as
\be
C\=g_0+3g_1\ ,
\ee
while the second equation in (\ref{w2}) determines the form of the 
bosonic potential,
\bea
V_B &=& 2 g_0^2\,(x^1{+}x^2{+}x^3{+}x^4)^{-2}\ +\ 2 g_1^2 
\!\!\!\sum_{j>i<k<l\atop k\ne j\ne l} (x^i{+}x^j{-}x^k{-}x^l)^{-2} \ ,\\[6pt]
O_2 &=& \ \ \sfrac12\,(x^1{+}x^2{+}x^3{+}x^4)^{-2}\ +\ \quad\sfrac12\,
\!\!\!\sum_{j>i<k<l\atop k\ne j\ne l} (x^i{+}x^j{-}x^k{-}x^l)^{-2} \ ,
\eea
in tune with the minimal expressions~(\ref{minpot}). 
Notice that $g_0$ and $g_1$ are independent and may be set to zero 
individually, but not their quantum corrections.
This model was also found in~\cite{wyl}.
\vspace{0.5cm}

\noindent
{\it 5.3. All $\cN{=}4$ three- and four-particle models
          based on $A_1\oplus G_2$ and $F_4$}\\
\noindent
Let us finally make a more systematic search for 
$\cN{=}4$ superconformal three- and four-particle models,
where the sums in (\ref{Fansatz}) run over particular positive root systems
and all coefficients are left open.
We adopt our second solution strategy and first solve the WDVV equations.
The resulting admissible values for the coefficients~$h_\m$ already define 
all $U=0$ models, since a vanishing $U$ solves the first equation 
in~(\ref{YW}) trivially. We shall encounter a free parameter~$t$ in the
allowed values~$h_\m(t)$, for special values of which it is possible to
turn on some $g_\m$ in~$U$, i.e.~find nontrivial solutions to the first 
equation in~(\ref{YW}). Motivated by the already known solutions, 
we allow any positive root from $A_1\oplus G_2$ in the $n{=}3$ case 
and from $F_4$ in the $n{=}4$ case. 
The result of a computer analysis is given below.

\begin{center}
\begin{tabular}{|crc||cc|cc|cc|}
\multicolumn{3}{l}{$A_1\oplus G_2$} & \multicolumn{2}{c}{model 1}
 & \multicolumn{2}{c}{model 2} & \multicolumn{2}{c}{model 3} \\
\hline
pos.~root $\vec n^\m$ & $\#$ & type & $g_\m$ & $h_\m$ 
 & $g_\m$ & $h_\m$ & $g_\m$ & $h_\m$\\
\hline & & & & & & & & \\[-10pt]
$(1,\ph1,\ph1)$ & 1 & -- & $\ti$ & $\sf13$      
 & $\ti$ & $\ph\sf13$  & $\ti$ & $\ph\sf13$ \\[4pt]
$(1,-1,\ph0)$   & 3 & S  & $0$   & $\sf13{-}3t$ 
 & $\ti$ & $\ph\sf12$  & $0$   & $-\sf16$   \\[4pt]
$(1,\ph1,-2)$   & 3 & L  & $0$ & $t$          
 & $0$   & $-\sf1{18}$ & $\ti$ & $\ph\sf16$ \\[4pt]
\hline
\end{tabular}

\bigskip

\begin{tabular}{|crc||cc|cc|cc|cc|}
\multicolumn{3}{l}{$F_4$} 
 & \multicolumn{2}{c}{model 1} & \multicolumn{2}{c}{model 2} 
 & \multicolumn{2}{c}{model 3} & \multicolumn{2}{c}{model 4} \\
\hline
pos.~root $\vec n^\m$ & $\#$ & type 
 & $g_\m$ & $h_\m$ & $g_\m$ & $h_\m$ & $g_\m$ & $h_\m$ & $g_\m$ & $h_\m$ \\
\hline & & & & & & & & & & \\[-10pt]
$(1,\ph1,\ph1,\ph1)$ & 1 & S 
 & $0$ & $\sf1{12}{-}2t$ & $0$ & $\sf14{-}6t$ & $0$ & $0$ & $0$ & $0$ \\[4pt]
$(1,\ph1,-1,-1)$     & 3 & S
 & $0$ & $\sf1{12}{-}2t$ & $0$ & $\sf14{-}6t$ & $0$ & $0$ & $0$ & $0$ \\[4pt]
$(1,\ph1,\ph1,-1)$   & 4 & S
 & $0$ & $\sf1{12}{-}2t$ & $0$ & $0$ & $0$ & $\sf14{-}6t$ & $0$ & $0$ \\[4pt]
$(2,\ph0,\ph0,\ph0)$ & 4 & S
 & $0$ & $\sf1{12}{-}2t$ & $0$ & $0$ & $0$ & $0$ & $0$ & $\sf14{-}6t$ \\[4pt]
$(2,\ph2,\ph0,\ph0)$ & 6 & L
 & $0$ & $t$ & $0$ & $t$ & $0$ & $t$ & $0$ & $t$ \\[4pt]
$(2,-2,\ph0,\ph0)$   & 6 & L
 & $0$ & $t$ & $0$ & $t$ & $0$ & $t$ & $0$ & $t$ \\[4pt]
\hline
\end{tabular}

\medskip

\begin{tabular}{|crc||cc|cc|cc|cc|}
\multicolumn{3}{l}{$F_4$ continued} 
 & \multicolumn{2}{c}{model 5} & \multicolumn{2}{c}{model 6} 
 & \multicolumn{2}{c}{model 7} & \multicolumn{2}{c}{model 8} \\
\hline
pos.~root $\vec n^\m$ & $\#$ & type
 & $g_\m$ & $\ \ h_\m\ \ \,$ & $g_\m$ & $\ \ h_\m\ \,$ 
 & $g_\m$ & $\ \ h_\m\ \,$ & $g_\m$ & $\ \ h_\m\ \,$\\
\hline & & & & & & & & & & \\[-10pt]
$(1,\ph1,\ph1,\ph1)$ & 1 & S
 &$\ti$& $\sf14$      &$\ti$& $\sf14$ & $0$ & $0$     & $0$ & $0$     \\[4pt]
$(1,\ph1,-1,-1)$     & 3 & S
 & $0$ & $\sf14{-}4t$ &$\ti$& $\sf14$ & $0$ & $0$     & $0$ & $0$     \\[4pt]
$(1,\ph1,\ph1,-1)$   & 4 & S
 & $0$ & $0$          & $0$ & $0$     &$\ti$& $\sf14$ & $0$ & $0$     \\[4pt]
$(2,\ph0,\ph0,\ph0)$ & 4 & S
 & $0$ & $0$          & $0$ & $0$     & $0$ & $0$     &$\ti$& $\sf14$ \\[4pt]
$(2,\ph2,\ph0,\ph0)$ & 6 & L
 & $0$ & $0$          & $0$ & $0$     & $0$ & $0$     & $0$ & $0$     \\[4pt]
$(2,-2,\ph0,\ph0)$   & 6 & L
 & $0$ & $t$          & $0$ & $0$     & $0$ & $0$     & $0$ & $0$     \\[4pt]
\hline
\end{tabular}
\end{center}

\noindent
In these tables, $\#$ is the number of positive roots obtained by
permuting the entries of the displayed vector, `type' refers to 
short (S) or long (L) roots, and $\ti$ indicates a free coupling~$g_\m$.
The free parameter~$t$ reflects the freedom of shifting the weights
between the short and the long roots.

For $n{=}3$, all models have translation-invariant relative motion, and all 
(except model~1 for $t{=}0$ and $t{=}\sfrac19$) 
exploit the full $G_2$ root system through~$F$. 
Model~1 has $U_{\rm rel}=0$, but models 2 and~3 with a nontrivial~$U_{\rm rel}$ 
arise at the special values of $t=-\sfrac1{18}$ and $t=\sfrac16$, respectively.
Model~2 was constructed in subsection~5.1, 
and all three indeed appear in~\cite{wyl}.

For $n{=}4$, only models 5 and~6 feature translation-invariant relative motion,
and only model~1 uses all roots of~$F_4$. 
Models 1 through~4 have $U=0$, and model~5 shows $U_{\rm rel}=0$, 
leaving models 6, 7 and 8 with a nontrivial~$U_{\rm rel}$.
The latter three arise at the special point $t=0$ of the corresponding models
listed above them. Models 1 to~4 all intersect at $t=\sfrac1{24}$, but
model~2 also agrees with model~5 at $t=0$ (where it becomes model~6).
Model~6 was presented in subsection~5.2 and also by Wyllard~\cite{wyl},
who insisted in relative translation invariance. Furthermore, it is interesting
to characterize the eight models (plus some special $t$ values) 
by the subalgebra of~$F_4$ each root system generates:

\begin{center}
\begin{tabular}{|c||cc|cc|cc|c|}
\hline
model number & {}\quad 1 \quad{} & $t{=}0,\sf1{24}$ 
& \ \ 2, 3, 4 \ \ & $t{=}\sf1{24}$ & 5 & $t{=}\sf1{16}$ & \ \ 6, 7, 8 \ \ \\
\hline & & & & & & & \\[-10pt]
$\#$ pos.\  roots & 24 & 12 & 16 & 12 & 10 &  7 &  4 \\[4pt]
dimension         & 52 & 28 & 36 & 28 & 24 & 18 & 12 \\[4pt]
subalgebra & $F_4$ & $D_4$ 
& $B_4$ & $D_4$ & $A_1{\oplus}B_3$ & $A_1{\oplus}A_3$ & $A_1^4$ \\[4pt]
\hline
\end{tabular}
\end{center}

\noindent
For the reader's convenience, 
we finally display the bosonic potentials for the models 5--8:
\be
\begin{aligned}
&V_5\=\frac{2\,\widetilde{g}_0^2}{(x_1{+}x_2{+}x_3{+}x_4)^2} \ + 
\sum_{\textrm{3 perms}}
\frac{\frac12(1{-}16\,t^2)^2\hbar^2}{(x_i{+}x_j{-}x_k{-}x_l)^2}\ +
\sum_{\textrm{6 perms}}
\frac{16\,t^2\hbar^2}{(x_i{-}x_j)^2}
\ +\ O(\psi^2,\lefteqn{\psi^4) \ ,}\\
&V_6\=\frac{2\,\widetilde{g}_0^2}{(x_1{+}x_2{+}x_3{+}x_4)^2} \ + 
\sum_{\textrm{3 perms}}
\frac{2\,\widetilde{g}_1^2}{(x_i{+}x_j{-}x_k{-}x_l)^2}
\ +\ O(\psi^2,\psi^4) \ ,\\
&V_7\=\sum_{\textrm{4 perms}}
\frac{2\,\widetilde{g}_2^2}{(x_i{+}x_j{+}x_k{-}x_l)^2}
\ +\ O(\psi^2,\psi^4) \ ,\\
&V_8\=\sum_{\textrm{4 perms}}
\frac{2\,\widetilde{g}_3^2}{x_i^2}
\ +\ O(\psi^2,\psi^4) \ ,
\end{aligned}
\ee
with $O(\psi^2,\psi^4)$ being Weyl ordered and  
$\ \widetilde{g}_\m^2=g_\m^2+\sfrac14\hbar^2$.
The central charge is $\ C=\sum_\m\#_\m g_\m$.
\vspace{0.8cm}

\noindent
{\bf 6. Conclusion}\\

\noindent
In this paper the transformation of generic conformal multi-particle mechanics
into a non-interacting system with nonlocal conformal symmetry~\cite{glp} was 
extended to accommodate $\cN{=}4$ supersymmetry. This step facilitates 
the construction of new $su(1,1|2)$ invariant many-body systems.
More concretely, for a potential ansatz quartic in the fermionic coordinates,
the closure of the superalgebra gave rise to a set of ``structure equations''
(\ref{w1}) and~(\ref{w2}) for two scalar (pre)potentials $U$ and~$F$ 
determining the potential~$V$, including quantum corrections.

For the $n$-body functions $U$ and $F$ we made an ansatz based on the choice
of a root system, with couplings $g$ and $h$, respectively, for each kind of
root. This reduced the structure equations to (\ref{YW}) with~(\ref{YWform}), 
i.e.~quadratic algebraic WDVV-type equations for~$\pa\pa\pa F$ and linear 
differential equations for~$\pa U$ in the $F$~background.
We fully analyzed these equations for the case of three and four particles
and found various solutions, based on the root systems of $A_1\oplus G_2$
and $F_4$, respectively. The $G_2$-type models are identical to those of
Wyllard~\cite{wyl}, whereas in the $F_4$~case we extend his result 
(our model~6) by several other solutions featuring translationally 
non-invariant relative particle motion.
Results based on higher-dimensional root systems will be reported elsewhere.

For three particles, the generality of our ansatz was proved by explicit
integration of the structure equations (\ref{w1}) and~(\ref{w2}).
With a growing number of particles, this becomes rather involved
because these equations are very rigid. For a general solution (unbiased
by the root-system ansatz) beyond $n{=}3$ a more advanced technique
is needed.

Turning to possible further developments, it would be interesting 
to generalize the present analysis to models exhibiting a 
$D(2,1|\alpha)$ symmetry and to the $\cN{=}8$ superconformal models 
constructed recently in~\cite{bikl,div}.
One may also attempt to construct an off-shell superfield description.
Finally, it is an open question whether the integrability of $\cN{=}4$
superconformal multi-particle models is tied to the root systems of
certain Lie superalgebras.
\vspace{0.5cm}

\noindent{\bf Acknowledgments}\\
\noindent
We are indebted to E.~Ivanov, S.~Krivonos, A.~Nersessian, F.~Toppan and
N.~Wyllard for useful discussions.
A.G.\ thanks the Institut f\"ur Theoretische Physik at
Leibniz Universit\"at Hannover for the hospitality extended to him at
different stages of this work.
We are grateful to the Joint Institute for Nuclear Research at Dubna
for providing a stimulating atmosphere during the workshop SQS~'07.
The research was supported by RF Presidential grants MD-8970.2006.2,
NS-4489.2006.2, INTAS grant 03-51-6346, DFG grant 436 RUS 113/669/0-3,
RFBR grant 06-02-16346, RFBR-DFG grant 06-02-04012 and the Dynasty Foundation.
\vspace{1cm}

\noindent
{\bf Appendix A}\\

\noindent
Given fermionic operators $\p_1,\dots,\p_n$, 
the Weyl-ordered product is defined as follows,
\begin{eqnarray*}
&&
\langle\p_1\p_2\rangle\=\sfrac{1}{2}\bigl(\p_1\p_2-\p_2\p_1\bigr)\ ,\\[2pt]
&&
\langle\p_1\p_2\p_3\rangle\=\sfrac{1}{3}\bigl(\p_1\langle\p_2\p_3\rangle +
\p_2\langle\p_3\p_1\rangle + \p_3\langle\p_1\p_2\rangle)\ ,\\[2pt]
&&
\langle\p_1\p_2\p_3\p_4\rangle\=\sfrac{1}{4}\bigl(
\p_1\langle\p_2\p_3\p_4\rangle - \p_2\langle\p_3\p_4\p_1\rangle +
\p_3\langle\p_4\p_1\p_2\rangle - \p_4\langle\p_1\p_2\p_3\rangle \bigr)
\end{eqnarray*}
etc., such that for any two neighboring operators one has
\begin{equation*}
\langle\dots\p_i\p_j\dots\rangle\=-\langle\dots\p_j\p_i\dots\rangle\ .
\end{equation*}
For deriving (\ref{restr}) it is convenient to pass from Weyl-ordered
operators to $qp$-ordered ones. In particular, for completely symmetric 
functions $M_{ij}$ and $L_{ijkl}$ one has
\begin{eqnarray*}
&&
M_{ij}\,\langle\p^i_\a \bar\p^{j\a}\rangle \=
M_{ij}\,\p^i_\a \bar\p^{j\a} - \hbar\,M_{kk}\ ,\\[2pt]
&&
W_{ijk}\,\langle\p^{i\b} \p^j_\b \bar\p^k_\a\rangle \= 
W_{ijk}\,\p^{i\b} \p^j_\b \bar\p^k_\a - \hbar\,W_{kki}\,\p^i_\a\ ,\\[2pt]
&&
W_{ijk}\,\langle\p^{i\a} \bar\p^j_\b \bar\p^{k\b}\rangle \=
W_{ijk}\,\p^{i\a} \bar\p^j_\b \bar\p^{k\b} + \hbar\,W_{kki}\,\bar\p^{i\a}
\ ,\\[2pt]
&&
L_{ijkl}\,\langle\p^{i\a} \p^j_\a \bar\p^k_\b \bar\p^{l\b}\rangle \=
L_{ijkl}\,\p^{i\a} \p^j_\a \bar\p^k_\b \bar\p^{l\b} - 
2\hbar\,L_{ijkk}\,\p^i_\a \bar\p^{j\a} + \hbar^2\,L_{kkpp}\ .
\end{eqnarray*}
\vspace{1cm}

\noindent
{\bf Appendix B}\\

\noindent
Here we present the explicit form of the WDVV coefficients for the 
three-body $\cN{=}4$ superconformal Calogero model~(\ref{fe}):
\begin{eqnarray*}
18 W_{112}&=&
+\frac{9}{x_1{-}x_2}-\frac{4}{2x_1{-}x_2{-}x_3}
+\frac{2}{2x_2{-}x_1{-}x_3}-\frac{1}{2x_3{-}x_1{-}x_2}
-\frac{6}{x_1{+}x_2{+}x_3}\ ,\\[2pt]
18 W_{113}&=&
+\frac{9}{x_1{-}x_3}-\frac{4}{2x_1{-}x_2{-}x_3}
-\frac{1}{2x_2{-}x_1{-}x_3}+\frac{2}{2x_3{-}x_1{-}x_2}
-\frac{6}{x_1{+}x_2{+}x_3}\ ,\\[2pt]
18 W_{122}&=&
-\frac{9}{x_1{-}x_2}+\frac{2}{2x_1{-}x_2{-}x_3}
-\frac{4}{2x_2{-}x_1{-}x_3}-\frac{1}{2x_3{-}x_1{-}x_2}
-\frac{6}{x_1{+}x_2{+}x_3}\ ,\\[2pt]
18 W_{123}&=&
+\frac{2}{2x_1{-}x_2{-}x_3}+\frac{2}{2x_2{-}x_1{-}x_3}
+\frac{2}{2x_3{-}x_1{-}x_2}-\frac{6}{x_1{+}x_2{+}x_3}\ ,\\[2pt]
18 W_{133}&=&
-\frac{9}{x_1{-}x_3}+\frac{2}{2x_1{-}x_2{-}x_3}
-\frac{1}{2x_2{-}x_1{-}x_3}-\frac{4}{2x_3{-}x_1{-}x_2}
-\frac{6}{x_1{+}x_2{+}x_3}\ ,\\[2pt]
18 W_{223}&=&
+\frac{9}{x_2{-}x_3}-\frac{1}{2x_1{-}x_2{-}x_3}
-\frac{4}{2x_2{-}x_1{-}x_3}+\frac{2}{2x_3{-}x_1{-}x_2}
-\frac{6}{x_1{+}x_2{+}x_3}\ ,\\[2pt]
18 W_{233}&=&
-\frac{9}{x_2{-}x_3}-\frac{1}{2x_1{-}x_2{-}x_3}
+\frac{2}{2x_2{-}x_1{-}x_3}-\frac{4}{2x_3{-}x_1{-}x_2}
-\frac{6}{x_1{+}x_2{+}x_3}\ ,\\[2pt]
18 W_{111}&=&
-\frac{9}{x_1{-}x_2}-\frac{9}{x_1{-}x_3}
+\frac{8}{2x_1{-}x_2{-}x_3}-\frac{1}{2x_2{-}x_1{-}x_3}
-\frac{1}{2x_3{-}x_1{-}x_2}-\frac{6}{x_1{+}x_2{+}x_3} ,\\[2pt]
18 W_{222}&=&
+\frac{9}{x_1{-}x_2}-\frac{9}{x_2{-}x_3}
-\frac{1}{2x_1{-}x_2{-}x_3}+\frac{8}{2x_2{-}x_1{-}x_3}
-\frac{1}{2x_3{-}x_1{-}x_2}-\frac{6}{x_1{+}x_2{+}x_3}\ ,\\[2pt]
18 W_{333}&=&
+\frac{9}{x_1{-}x_3}+\frac{9}{x_2{-}x_3}
-\frac{1}{2x_1{-}x_2{-}x_3}-\frac{1}{2x_2{-}x_1{-}x_3}
+\frac{8}{2x_3{-}x_1{-}x_2}-\frac{6}{x_1{+}x_2{+}x_3}\ .
\end{eqnarray*}
The quantum correction 
$\hbar^2 O_2=\sfrac{\hbar^2}{8} W_{ijk}W_{ijk}$ 
to the two-body Calogero potential was given in~(\ref{n3qu})
and involves three-body interactions.

\end{document}